\documentclass[11pt]{article}
\usepackage[english]{babel}
\input{epsf}
\usepackage{mathtext}
\usepackage{graphics,dcolumn,bm,float}
\usepackage{amssymb,amsmath,rotate,color}
\usepackage{times}
\usepackage{amsmath}
\usepackage{amsfonts}
\usepackage{amssymb}
\begin{document}
\title{Dirac operator on the quantum fuzzy four-sphere $ S_{qF}^{4} $}
\author{M. Lotfizadeh\thanks{E-mail: M. Lotfizadeh@urmia.ac.ir}$^{\;\;,1}$  \\
\footnotesize\textit{$^1$Department of Physics, Faculty of Science, Urmia University, P.O.Box: 165, Urmia, Iran}\\
\footnotesize\textit{}}
\date{}
\maketitle
\begin{abstract}
The q-deformed fuzzy Dirac and chirality operators on quantum fuzzy four-sphere $ S^{4}_{qF} $. Using the q-deformed fuzzy Ginsparg-Wilson algebra, it has been studied the q-deformed fuzzy Dirac and chirality operators in instanton and no-instanton sector. Also, gauged Dirac and chirality operators in both cases have been constructed. It has been showed that In each step our results have correct commutative limit in the limit case when $ q\longrightarrow1 $ and noncommutative parameter $ l $ tends to infinity. 
\end{abstract}

\section{Introduction}

Dirac and chirality operators are two important self-adjoint operators for the Connes-Lott approach to noncommutative geometry. A unital spectral triple [1-2], $ (\mathcal{A},\mathcal{H},D) $ consists of a complex unital $ \ast $-algebra $ \mathcal{A} $, faithfully $ \ast $-represented by bounded operators on a separable Hilbert space $\mathcal{H} $, and a self-adjoint operator $ D: \mathcal{H}\rightarrow \mathcal{H} $ (the Dirac operator) with the following properties:

$ \bullet $the resolvent $ (D-\lambda)^{-1},\,\lambda \notin \mathbb{R} $ is a compact operator on $\mathcal{H} $, 

$ \bullet $for all $ a \in \mathcal{A} $ the commutator $ [D, \pi(a)] $ is a bounded operator on $\mathcal{H}$.

A spectral triple $ (\mathcal{A},\mathcal{H},D) $ is called even if there exists a $ \mathbb{Z}_{2} $-grading of $\mathcal{H}$, i.e. An operator $ \gamma :\mathcal{H}\rightarrow \mathcal{H}$ with $ \gamma^{\ast}=\gamma $ and $ \gamma^{2}=1$, such that $ \gamma D+D\gamma=0 $ and $ \gamma a=a \gamma $ for all $ a \in \mathcal{A} $. Otherwise the spectral triple is said to be odd. For odd dimensional manifolds, there are no chirality operators and in such a case, the Dirac operator describes only the differential structures. There are three types of Dirac and chirality operators on the fuzzy two-sphere. Ginsparg-Wilson Dirac operator, $ D_{GW} $ [3-10], Watamura-Watamura Dirac operator $ D_{WW} $ [11-13] and Grosse-Klimcik-Presnajder Dirac operator $ D_{GKP} $ [14-15]. These three types of Dirac operators are compared with each other in [16].
The idea of q-deformed geometry was extensively studied in the late 1980 s and 1990 s. The q-deformed Hopf fibration has been studied in the frame work of Hopf-Galois extension in [17]. Podles sphere are introduced in [18-19]. The q-deformed Dirac operator on quantum Podles sphere has been studied from different approaches [20-25].\\ In contrast to the noncommutative two-sphere and Dirac operator on it, which has been widely studied, few references have been studied the Dirac operator on noncommutative four-sphere. Physics on fuzzy four-sphere $ S_{F}^{4} $ has been studied in [28-35]. In[28], the truncated fuzzy four-sphere and the action of scalar field on it are studied. Longitudinal five-branes as fuzzy four-spheres in matrix theory is studied in [29]. In [30], spherical harmonics for fuzzy four-sphere is constructed. In [31], using a matrix model with a fifth-rank Chern-Simons term, a noncommutative gauge theory over a fuzzy 4-sphere is studied. Scalar field theory on $ S_{F}^{4} $ has been studied in [32]. Projective module description of fuzzy four-sphere has been investigated in [33,34]. In[35], generalized 4-dimensional fuzzy spheres with twisted extra dimensions has been studied. These spheres can be viewed as $SO(5)-$ equivariant projections of quantized coadjoint orbits of $SO(6) $. It has been showed that they arise as solutions in Yang- Mills matrix models, which naturally leads to higher-spin gauge theories on $ S^{4} $. \\ The quantum four-sphere $ S_{q}^{4} $ as the base manifold of the quantum version of second Hopf bundle has been constructed in [36-39]. The quantum group $ SO_{q}(5) $ and the representation of its corresponding algebra $ \mathcal{U}_{q}(so(5)) $, as the symmetry group of $ S_{q}^{4} $, are studied in [41-43]. \\
 In this paper our aim is to study the gauged q-deformed  fuzzy Ginsparg-Wilson Dirac and chirality operators in instanton  and no-instanton sector using the gauged q-deformed fuzzy Ginsparg-Wilson algebra on quantum fuzzy four- sphere $ S_{qF}^{4} $. \\
 This paper is organized as follows:
 In section 2, we review the groups $ SO(5) $ and $ SO_{q}(5) $ as symmetry groups of $ S^{4} $ and $ S_{qF}^{4} $. In section 3 we briefly review the 4-sphere $ S^{4} $, fuzzy 4-sphere $ S_{F}^{4} $ and q-deformed fuzzy 4-sphere $ S_{qF}^{4} $. In section 4 we briefly review the second Hopf fibration $ S^{7}\longrightarrow S^{4} $ with special focus on projectors and projective modules of the sections of this principal bundle. Also, we study the q-deformed fuzzy version of this fibration.
Specially we study the q-deformed projectors and q-deformed projective modules of the Hopf-Galois extension of the fibration $ S^{7}\longrightarrow S^{4} $.In section 5  Spin $\frac{1}{2}$ q-deformed projectors of the left and right q-deformed projective $ \mathcal{A}(S_{qF}^{4}) -$module have been studied. The q-deformed fuzzy Ginsparg-Wilson algebra has been constructed in section 6. Also, in this section we have been constructed q-deformed fuzzy chirality and Dirac operators using the q-deformed left and right projectors and their corresponding idempotents. In section 7, gauged q-deformed fuzzy Dirac operator has been studied. In section 8 it has been constructed the q-deformed fuzzy Dirac operator in instanton sector. Finally, in section 9, gauging the q-deformed fuzzy Dirac operator in instanton sector has been studied. In each step, we compare our results with the limit case $ q\longrightarrow 1 $ and $ l\longrightarrow \infty $ .

\section{$ so(5) $ and $\mathcal{U}_{q}(so(5))$ }

Let us denote by $ \sigma_{1}, \sigma_{2} $ and $ \sigma_{3} $ Pauli matrices, which are the generators of the $ su(2) $ Lie algebra. They also, span the Clifford algebra Cliff(3): $ \lbrace \sigma_{i}, \sigma_{j}\rbrace = i\delta_{ij}\mathbf{1} $. Using Pauli matrices, one can construct $ 4\times 4 $ matrices $ G_{i}, i=1,2,...,5 $ as follow:
\begin{equation}
G_{j}= \begin{pmatrix}
0   &   -i\sigma_{j}\\
i\sigma_{j}  &  0
\end{pmatrix}, \quad j=1,2,3,\quad G_{4}= \begin{pmatrix}
0   &   \mathbf{1}_{2}\\
\mathbf{1}_{2}  &  0
\end{pmatrix},\quad G_{5}= \begin{pmatrix}
\mathbf{1}_{2}  &   0\\
0  &  -\mathbf{1}_{2}
\end{pmatrix}\tag{2-1}.
\end{equation}
The matrices $ G_{i}(i=1,...,4) $ and $ G_{i}(i=1,...,5) $ span Clifford algebra  Cliff(4) and Cliff(5), respectively: $ \lbrace G_{i}, G_{j} \rbrace =2\delta_{ij}\mathbf{1}. $ They also satisfy the following relations
\begin{equation}
G_{i}^{2}= \mathbf{1}, \quad \varepsilon^{ijkm}G_{i}G_{j}G_{k}G_{m}=24 G_{m}\tag{2-2}.
\end{equation}
Note that $ SO(5) $ is a compact simple Lie group of rank two and the irreducible representations are specified by two indexes $ m_{1} $ and $ m_{2} $. 
 The matrices (2-1) are the $ (1/2,1/2) $ representation of $ Spin(5) $: the double covering of $ SO(5) $. The matrices $ G_{i} $ are not a basis for all matrices (they are not closed). To construct a basis, we need to include the commutators of these matrices, too. So, let us define:
\begin{equation}
G_{ij}:= \dfrac{1}{2i•}[G_{i}, G_{j}], \quad i\neq j, \tag{2-3}
\end{equation}
which satisfy the Lie algebra of $ SO(5) $ Lie group. These matrices are $ (1/2,1/2) $ spin representation of the double covering of this algebra, i.e. $ spin(5) $ or equivalently $ sp(2) $. To construct the $ (l/2,l/2) $ representation of this algebra, let us construct the $ l $-fold symmetrized tensor product of the $ (1/2,1/2) $ representation:  
\begin{equation}
G_{i}^{(l)}= (G_{i}\otimes 1\otimes ...\otimes 1+ 1\otimes G_{i}\otimes 1\otimes ... \otimes 1+ 1\otimes 1\otimes ... \otimes G_{i})_{sym},\tag{2-4}
\end{equation}
which generalize $ G_{i} $ to the l-th level and has the dimension of representation is $ d_{l}= \dfrac{1}{6•}(l+1)(l+2)(l+3) $. The Lie group $ SO(5) $ has the following Casimir:
\begin{equation}
C_{so(5)}= G_{i}^{(l)}G_{i}^{(l)}= l(l+4)\mathbf{1}.\tag{2-5}
\end{equation}
In the Lie group $ Spin(5) $, an arbitrary matrix can be considered as an element of the vector space $ (l/2,l/2)\otimes (\bar{l/2},\bar{l/2}) $ which reduces under $ Spin(5) $ as follow:
\begin{equation}
(l/2,l/2)\otimes (l/2,l/2)=\bigoplus_{m_{1}=0}^{l}\bigoplus_{m_{2}=0}^{m_{1}}(m_{1},m_{2}).\tag{2-6}
\end{equation}
where the dimension of $ (m_{1},m_{2}) $  is as follow[32]:
\begin{equation}
d_{(m_{1},m_{2})}= \dfrac{1}{6•}(2m_{1}+3)(2m_{2}+1)((m_{1}+1)(m_{1}+2)-m_{2}(m_{2}+1))\tag{2-7}
\end{equation}
For the special case $ l=1 $ the dimension of representation is $ 4 $.\\
Now, we can define the coordinates of $ S_{F}^{4} $ as follow:
\begin{equation}
X_{i}= \dfrac{G_{i}^{(l)}}{\sqrt{l(l+3)}•}, \quad X_{i}X_{i}= 1,\quad i=1,...,5.\tag{2-8}
\end{equation}
When the parameter $ l $ tends to infinity, these operators construct the commutative algebra $ C^{\infty}(S^{4}) $ with $ [X_{i}, X_{j}]=0. $ The generators $ X_{i} $ does not construct a closed algebra. To solve this problem we have to add some generators to $ X_{i}^{,} $s. Similar to $ (1/2,1/2) $ representation, let us define the generators $ G_{ij}^{(l)} $ as follow:
\begin{equation}
G_{ij}^{(l)}:= \dfrac{1}{2•}[G_{i}^{(l)}, G_{j}^{(l)}], \quad G_{ij}^{(l)}G_{ij}^{(l)}= 4l(l+4)\mathbf{1}.\tag{2-9}
\end{equation}
Now, we can define the new coordinates as follow:
\begin{equation}
X_{ij}= \dfrac{G_{ij}^{(l)}}{•\sqrt{l(l+3)}}\tag{2-10}
\end{equation}
this shows that the noncommutativity is as a result of the presence of the extra coordinates $ X_{ij}. $ The authors in [44] showed that the fuzzy four-sphere is given by the coset $ \dfrac{SO(5)}{U(2)•}, $ which shows that a fuzzy two-sphere is attached to each point of the $ S_{F}^{4}. $ So the little group of $ S_{F}^{4} $ is not $ SO(4) $ but $ U(2). $

Noncommutative geometry is a pointless geometry. In this geometry instead of the coordinates $ x_{i} $'s of $ S^{4} $, the $ SO(5) $ generators in the unitary  irreducible $ (l/2,l/2) -$representation space have the role of the points of the fuzzy $ S^{4} $. Let us denote the generators of $ SO(5) $ by $ \mathbf{G}^{(l)}=(G_{i}^{(l)}, G_{ij}^{(l)}) $, and the fuzzy Hermitian matrix algebra by $\mathcal{A}_{l}=\{ \alpha\in Mat_{d_{l}} (\mathbb{C}) \}$. Every arbitrary element $ \alpha $ can be expressed in terms of the bases, as a unitary $ (l/2,l/2)- $representation of the $ so(5) $, of the generator $\mathbf{G}^{(l)} $.\\
In the Hopf fibration  $S^{7}\;\xrightarrow{SU(2)}\;S^{4}$, the module of sections is $C(S^{4})$-module $\Gamma^{\infty}(S^{4},\mathcal{E}^{(n)})$ in which $C(S^{4})$ is the commutative algebra of functions on $S^{4}$. In the fuzzy case, this algebra is a noncommutative algebra and therefore, left and right modules are not isomorphic. In this case to each generator $\mathbf{G}^{(l)}$, we associate two linear operators $\mathbf{G}^{(l)L}$ and $\mathbf{G}^{(l)R}$ with the left and right actions on $\mathcal{A}_{l} $:
\begin{equation*}
G_{i}^{(l)L} \alpha=G_{i}^{(l)} \alpha ,\quad G_{i}^{(l)R} \alpha=\alpha G_{i}^{(l)} ,
\end{equation*}
\begin{equation}
G_{ij}^{(l)L} \alpha=G_{ij}^{(l)} \alpha ,\quad G_{ij}^{(l)R} \alpha=\alpha G_{ij}^{(l)} ,\ \quad\forall\alpha \in \mathcal{A}_{l},
\tag{2-11}
\end{equation}
these left and right operators commute with each other:
\begin{equation}
[G_{i}^{(l)L},G_{j}^{(l)R}]=0,\quad [G_{ij}^{(l)L},G_{km}^{(l)R}]=0
\tag{2-12}
\end{equation}
The $ G_{i}^{(l)L} $ and $ G_{j}^{(l)R} $ have the same $ so(5) $ algebra with the Casimir $ C_{so(5)} $:
\begin{equation}
C_{so(5)}= G_{i}^{(l)L}G_{i}^{(l)L}= G_{i}^{(l)R}G_{i}^{(l)R}=l(l+4)
\tag{2-13}
\end{equation}
\begin{equation}
C_{so(5)}= G_{ij}^{(l)L}G_{ij}^{(l)L}= G_{ij}^{(l)R}G_{ij}^{(l)R}=4l(l+4).
\tag{2-14}
\end{equation}
We use $ \mathbf{G}^{L} $, $ \mathbf{G}^{R} $ to define the fuzzy version of orbital momentum operator $\boldsymbol{\mathcal{G}}$ on the fuzzy $ S^{4}$. We define $\boldsymbol{\mathcal{G}}$ by the adjoint action of $ (G_{i}^{(l)},G_{ij}^{(l)}) $ on the $\mathcal{A}_{l}$:
\begin{equation}
\mathcal{G}_{i}^{(l)} \alpha =(G_{i}^{(l)L} - G_{i}^{(l)R})\alpha= ad_{G_{i}^{(l)}}\alpha = [G_{i}^{(l)},\alpha]\rightarrow -2i(X_{ij}\dfrac{\partial}{\partial X_{j}•}- X_{j}\dfrac{\partial}{\partial X_{ij}•})
\tag{2-15}
\end{equation}
\begin{equation*}
\mathcal{G}_{ij}^{(l)} \alpha =(G_{ij}^{(l)L} - G_{ij}^{(l)R})\alpha= ad_{G_{ij}^{(l)}}\alpha =
\end{equation*}
\begin{equation}
 [G_{ij}^{(l)},\alpha]\rightarrow 2(X_{i}\dfrac{\partial}{\partial X_{j}•}- X_{j}\dfrac{\partial}{\partial X_{i}•}-X_{ik}\dfrac{\partial}{\partial X_{kj}•}+ X_{jk}\dfrac{\partial}{\partial X_{ki}•}  )
\tag{2-16}
\end{equation}
We define the quantum numbers $ [n]_{q} $ and $ [n]_{q^{2}} $ as follow:
\begin{equation}
[n]_{q}= \dfrac{q^{n}- q^{-n}}{q- q^{-1}•}, \quad [n]_{q^{2}}= \dfrac{q^{2n}- q^{-2n}}{q^{2}- q^{-2}•}\tag{2-17}
\end{equation}

Now, for $ 0< q<1 $ let us construct the real form of the Drinfeld-Jimbo deformation of $ so_{\mathbb{C}}(5) $, it is a real form of the Hopf algebra $ \widehat{\mathcal{U}}_{q}(so(5)) $[38]. As a $ *- $algebra, $ \mathcal{U}_{q}(so(5))\cong so_{q}(5) $ is generated by the Chevally generators consists of two triplets $ (J_{\pm 1}, J_{+0}) $ and $ (J_{\pm 2}, J_{-0}) $ corresponding to the roots $ 1 $ and $ 2 $, respectively. The $ \mathcal{U}_{q}(so(5)) $ Hopf algebra can be given by the following relations[41]:
\begin{equation*}
[J_{+1}, J_{-1}]= \dfrac{J_{+0}^{2}- J_{+0}^{-2}}{q-q^{-1}•}, \quad
[J_{+2}, J_{-2}]= \dfrac{J_{-0}^{2}- J_{-0}^{-2}}{q^{2}-q^{-2}•}
\end{equation*}
\begin{equation*}
[J_{+1}, J_{-2}]= [J_{+2}, J_{-1}]=0, \quad J_{\pm 1}^{*}= J_{\mp 1}, J_{\pm 2}^{*}= J_{\mp 2},
\end{equation*}
\begin{equation*}
J_{+0}J_{\pm1}= q^{\pm 1}J_{\pm 1}J_{+0},\quad  J_{+0}J_{\pm 2}= q^{\mp 1}J_{\pm 2}J_{+0}
\end{equation*}
\begin{equation*}
J_{-0}J_{\pm2}= q^{\pm 1}J_{\pm 2}J_{-0},\quad  J_{-0}J_{\pm 1}= q^{\mp 1}J_{\pm 1}J_{-0}
\end{equation*}
\begin{equation*}
J_{\pm1}^{3}J_{\pm2}- (q^{2}+q^{-2}+1)J_{\pm1}^{2}J_{\pm2}J_{\pm1}+ (q^{2}+q^{-2}+1)J_{\pm1}J_{\pm2}J_{\pm1}^{2}- J_{\pm2}J_{\pm1}^{3}=0,
\end{equation*}
\begin{equation}
J_{\pm2}^{2}J_{\pm1}- (q^{2}+q^{-2})J_{\pm2}J_{\pm1}J_{\pm2}+ J_{\pm1}J_{\pm2}^{2}=0,\tag{2-18}
\end{equation}
The coproduct $ \Delta: \mathcal{U}_{q}(so(5))\rightarrow \mathcal{U}_{q}(so(5))\otimes \mathcal{U}_{q}(so(5)) $, counit $ \varepsilon: \mathcal{U}_{q}(so(5))\rightarrow \mathbb{C} $ and antipode $ S: \mathcal{U}_{q}(so(5))\rightarrow \mathcal{•U}_{q}(so(5)) $
 of the Hopf algebra $ \mathcal{U}_{q}(so(5)) $ are given by:
\begin{equation*}
\Delta J_{\pm 0}= J_{\pm 0}\otimes J_{\pm 0},
\end{equation*}
\begin{equation*}
\Delta J_{\pm i}= J_{\pm i}\otimes J_{\pm 0}+ J_{\pm 0}\otimes J_{\pm i}, \quad i=1,2,
\end{equation*}
\begin{equation*}
\varepsilon(J_{\pm 0})=1,\quad \varepsilon(J_{\pm i})= 0, \quad i=1,2,
\end{equation*}
\begin{equation}
S(J_{\pm 0})= J_{\pm 0}^{-1},\quad S(J_{\pm i})= -q^{i}J_{\pm i}, \quad i=1,2,\tag{2-19}
\end{equation}
The algebra $ \mathcal{A}(S_{q}^{4}) $ is an $ \mathcal{U}_{q}(so(5))- $module $ *- $algebra for the action given by:
\begin{equation*}
J_{\pm 0}\triangleright x_{i}= qx_{i}, \quad i=1,2, \quad J_{-0}\triangleright x_{1}= q^{-1}x_{1},
\end{equation*}
\begin{equation*}
J_{+1}\triangleright x_{0}= q^{-1/2}x_{1}, \quad J_{+2}\triangleright x_{1}= x_{2},\quad J_{-1}\triangleright x_{1}= q^{1/2}[2]_{q}x_{0},
\end{equation*}
\begin{equation}
J_{-1}\triangleright x_{0}= -q^{-3/2}x_{1}^{*},\quad J_{-2}\triangleright x_{2}= x_{1},\tag{2-20}
\end{equation}
and
\begin{equation}
J_{\pm 0}\triangleright x_{j}= x_{j},\quad J_{+i}\triangleright x_{j}= 0,\quad J_{-i}\triangleright x_{j}=0,\tag{2-21}
\end{equation}
in all other cases.
For each nonnegative $ m_{1} $ and $ m_{2} $ such that $ m_{2}\in \dfrac{1}{2•}\mathbb{N}$ and $ m_{2}-m_{1}\in \mathbb{N} $, there is an irreducible representation of $ \mathcal{U}_{q}(so(5)) $ whose representation space is $ \mathcal{V}_{(m_{1},m_{2})} $. The highest weight irreducible representations of $ \mathcal{U}_{q}(so(5)) $ are $ (0,l) $ and $ (1/2,l) $. Let us use the notation $ \mathcal{V}_{l}:= \mathcal{V}_{(0,l)} $ if $ l \in \mathbb{N} $ and $ \mathcal{V}_{l}:= \mathcal{V}_{(1/2,l)} $ if $ l \in \mathbb{N}+1/2 $. The vector space $ \mathcal{V}_{l} $, for all $ l \in 1/2 \mathbb{N} $ has orthonormal bases $ |l,m_{1},m_{2},j> $ where for a fix $ l $ we have two cases:\\
\begin{equation}
  \quad l\in \mathbb{N} \quad j=0,1,...,l ,\quad j-|m_{1}|\in \mathbb{N}, \quad l-j-|m_{2}|\in 2\mathbb{N},\tag{2-22}
 \end{equation}
 \begin{equation}
  \quad l\in \mathbb{N}+ \dfrac{1}{2•} \quad j=1/2,3/2,...,l-1,l ,\quad j-|m_{1}|\in \mathbb{N}, \quad l+1/2-j-|m_{2}|\in \mathbb{N},\tag{2-23}
\end{equation}
It has been showed that there is an isomorphism $ \mathcal{A}(S_{q}^{4})\cong \oplus_{l\in \mathbb{N}}\mathcal{V}_{l} $ of $ \mathcal{U}_{q}(so(5)) $ left modules. In the representation space $ |l,m_{1},m_{2},j> $ the eigenvalue of the Casimir operator of $ \mathcal{U}_{q}(so(5)) $ is given as follow[42]:
\begin{equation}
C_{\mathcal{U}_{q}(so(5))}= ([m_{1}]_{q}[m_{1}+3]_{q}+ [m_{2}]_{q}[m_{2}+1]_{q}\dfrac{[2m_{1}+3]_{q^{2}}}{[2m_{1}+3]_{q}•}).\tag{2-24}
\end{equation}
In some special cases, the Casimir (2-24) reduces to the following Casimirs
\begin{equation}
m_{1}=l, \quad m_{2}=0, \quad C_{\mathcal{U}_{q}(so(5))}= [l]_{q}[l+3]_{q}\tag{2-25}
\end{equation}
\begin{equation}
m_{1}= m_{2}= l, \quad C_{\mathcal{U}_{q}(so(5))}= [l]_{q^{2}}[l+2]_{q^{2}}\tag{2-26}
\end{equation}
In the limit case when the quantum parameter q tends to unit, the Casimir () reduces to the following Casimir of $ SO(5) $
\begin{equation}
C_{so(5)}= (m_{1}(m_{1}+3)+ m_{2}(m_{2}+1))\tag{2-27}
\end{equation}
In the special case of $ (l/2,l/2) $ representation the Casimir (2-24) reduces to the following Casimir
\begin{equation}
C_{(l/2,l/2)}=l(l+4).\tag{2-28}
\end{equation}
For later use, let us define $\sigma: \mathcal{U}_{q}(so(5))\rightarrow Mat_{4}(\mathbb{C})  $ as the spin $ (1/2,1/2) *- $representation of the $ \mathcal{U}_{q}(so(5)) $ by the following matrices[Landi]:
\begin{equation*}
\sigma(J_{+0})=\Sigma_{+0}= \begin{pmatrix}
	q^{1/2} & 0 &0 & 0\\
	0 & q^{1/2} & 0 & 0\\
	0& 0& q^{-1/2} &0\\
	0& 0& 0& q^{-1/2}
	\end{pmatrix},\quad 
	\sigma(J_{-0})=\Sigma_{-0}= \begin{pmatrix}
	1 & 0 &0 & 0\\
	0 & q^{-1} & 0 & 0\\
	0& 0& q &0\\
	0& 0& 0& 1
	\end{pmatrix},\quad 	
	\end{equation*}
	\begin{equation}
	\sigma(J_{+1})=\Sigma_{+1}= \begin{pmatrix}
	0 & 0 &1 & 0\\
	0 & 0 & 0 & 1\\
	0& 0& 0 &0\\
	0& 0& 0& 0
	\end{pmatrix},\quad 
	\sigma(J_{+2})=\Sigma_{+2}= \begin{pmatrix}
	0 & 0 &0 & 0\\
	0 & 0 & 0 & 0\\
	0& 1& 0 &0\\
	0& 0& 0& 0
	\end{pmatrix}\
	\tag{2-29}
\end{equation}

\section{Four-sphere $ S^{4} $, fuzzy four-sphere $ S_{F}^{4} $ and q-deformed fuzzy four-sphere $ S_{qF}^{4} $}

As a subset of $ \mathbb{R}^{5} $, the four-dimensional sphere $ S^{4} $ is defined as:
\begin{equation}
S^{4}= \lbrace (x^{1}, x^{2},x^{3},x^{4},x^{5})\in \mathbb{R}^{5}, \quad \sum_{i=1}^{5} x^{i}x^{i}=1\rbrace,\tag{3-1}
\end{equation} 
and we let $ U\subset S^{4} $ denote the chart of $ S^{4} $ given by
\begin{equation*}
x^{1}= cos\xi_{1}cos\varphi cos\psi, \quad x^{2}= sin\xi_{1}cos\varphi cos\psi,\quad x^{3}= cos\xi_{2}sin\varphi cos\psi,
\end{equation*}
\begin{equation}
x^{4}= sin\xi_{2}sin\varphi cos\psi, \quad x^{5}= sin\psi,\tag{3-2}
\end{equation}
where $ 0<\xi_{1},\xi_{2}<2\pi, 0<\varphi<\pi/2, -\pi/2<\psi<\pi/2 $. Now, let us define 
\begin{equation}
x_{\pm1}=x^{1}\pm i x^{2}, \quad x_{\pm2}= x^{3}\pm ix^{4}, \quad x_{0}= x^{5},\tag{3-3}
\end{equation}
then, we have 
\begin{equation}
x_{\pm1}= e^{\pm i\xi_{1}}cos\varphi cos\psi, \quad x_{\pm2}= e^{\pm i\xi_{2}}sin\varphi cos\psi,\quad x_{0= sin\psi},\tag{3-4}
\end{equation}
It is easy to see that
\begin{equation}
\mathbf{x}\cdot \mathbf{x}= \sum_{i=1}^{5}x^{i}x^{i}= |x_{+1}|^{2}+ |x_{+2}|^{2}+ x_{0}^{2}= x_{+1}x_{-1}+x_{+2}x_{-2}+x_{0}^{2}=1.\tag{3-5}
\end{equation}
The standard four-sphere is a homogeneous space of the five-dimensional rotation group 
$ S^{4}\cong \dfrac{SO(5)}{SO(4)•}\cong \dfrac{Spin(5)}{Spin(4)•} $ and then it is a spin manifold but it is not a Kahler manifold that accommodate symplectic structure.
 
Noncommutative geometry is a pointless geometry. In this geometry instead of the coordinates $ x^{i} $ of $ S^{4} $, the $ so(5) $ generators in the unitary irreducible $ l- $representation space have the role of the points of the fuzzy four-sphere $ S_{F}^{4} $. Let $ G_{\pm1}^{(l)}, G_{\pm2}^{(l)}, G_{0}^{(l)}$ be the generators of $ so(5) $ constructed out of $ G_{i}^{(l),}s $:
\begin{equation}
G_{\pm1}^{(l)}= \dfrac{1}{2•}(G_{1}^{(l)}\pm iG_{2}^{(l)}), \quad G_{\pm2}^{(l)}= \dfrac{1}{2•}(G_{3}^{(l)}\pm iG_{4}^{(l)}), \quad G_{0}^{(l)}= G_{5}^{(l)}.\tag{3-6}
\end{equation}
 Between $ (m_{1},m_{2}) $ representations of $ so(5) $, only the representation $ (m_{1}=l,m_{2}=0) $ with the Casimir $ l(l+3) $ correspond to functions on $ S_{F}^{4} $, all others are non$ -S_{F}^{4} $ representations[32].
Then we define the coordinates of fuzzy four-sphere $ S_{F}^{4} $ as follow:
\begin{equation}
X_{\pm1}= \dfrac{G_{\pm1}^{(l)}}{\sqrt{l(l+3)}•}, \quad X_{\pm2}= \dfrac{G_{\pm2}^{(l)}}{\sqrt{l(l+3)}•},\quad X_{0}= \dfrac{G_{0}^{(l)}}{\sqrt{l(l+3)}•}.\tag{3-7}
\end{equation}
Now, let us introduce the noncommutative coordinates of $ S_{F}^{4} $ as follow:
\begin{equation*}
X^{1}= \dfrac{1}{\sqrt{2}•}(X_{+1}+X_{-1}),\quad X^{2}= \dfrac{1}{i\sqrt{2}•}(X_{+1}-X_{-1}),\quad X^{3}= \dfrac{1}{\sqrt{2}•}(X_{+2}+X_{-2}),
\end{equation*}
\begin{equation}
X^{4}= \dfrac{1}{i\sqrt{2}•}(X_{+2}-X_{-2}),\quad X^{5}= X_{0}.\tag{3-8}
\end{equation}
It is easy to see that
\begin{equation}
\mathbf{X}\cdot \mathbf{X}=\sum_{i=1}^{5}X^{i}X^{i}=(X_{+1}X_{-1}+ X_{-1}X_{+1})+(X_{+2}X_{-2}+ X_{-2}X_{+2})+X_{0}^{2}=1.\tag{3-9}
\end{equation}
In the limit case when the noncommutative parameter $ l $ tends to infinity these coordinates tend to commutative coordinates of $ S^{4} $:
\begin{equation}
\lim_{l \to \infty}\dfrac{X^{i}}{\sqrt{l(l+3)}•}= \lim_{l \to \infty}\dfrac{X^{i}}{l•}= x^{i}.\tag{3-10}
\end{equation}
It is important to notice, unlike the case of fuzzy two-sphere coordinates do not satisfy a closed algebra by themselves
\begin{equation}
[X^{i}, X^{j}]= \dfrac{X^{ij}}{\sqrt{l(l+3)}•}.\tag{3-11}
\end{equation}
With the $ SO(5) $ generators $ X_{ij} $ the fuzzy coordinates satisfy the following closed algebra:
\begin{equation*}
[X_{i}, X_{j}]= \dfrac{X_{ij}}{\sqrt{l(l+3)}•}, \quad [X_{i}, X_{jk}]= \dfrac{-i}{\sqrt{l(l+3)}•}(\delta_{ij}X_{k}-\delta_{ik}X_{j}),
\end{equation*}
\begin{equation}
[X_{ij}, X_{km}]= \dfrac{i}{\sqrt{l(l+3)}•}(\delta_{ik}X_{jm}-\delta_{im}X_{jk}+ \delta_{jk}X_{im}- \delta_{jm}X_{ik}),\tag{3-12}
\end{equation}
By identifying $ X_{i6}=1/2 X_{i} $, we find the above algebra is expressed by the $ SO(6) $ algebra
\begin{equation}
[X_{ab}, X_{cd}]= \dfrac{i}{\sqrt{l(l+3)}•}(\delta_{ac}X_{bd}- \delta_{ad}X_{bc}+ \delta_{bc}X_{ad}- \delta_{bd}X_{ac}), \quad a,b=1,2,...,6.\tag{3-13}
\end{equation}
Now, let us define the coordinates of quantum fuzzy four-sphere $ S_{qF}^{4} $ using the generators of $ \mathcal{U}_{q}(so(5)) $ as follow:
\begin{equation}
X_{\pm1}^{q}= \dfrac{J_{\pm1}}{\sqrt{[l]_{q}[l+3]_{q}}•}, \quad X_{\pm2}^{q}= \dfrac{J_{\pm2}}{\sqrt{[l]_{q}[l+3]_{q}}•}, \quad X_{0}^{q}= \dfrac{J_{0}}{\sqrt{[l]_{q}[l+3]_{q}}•}.\tag{3-14}
\end{equation}
In the limit case when $ q $ tends to unit the above equations reduce to (3-7), as well as in the limit cases when $ q\rightarrow 1 $ and $ l\rightarrow \infty $ equation (3-14) reduces to (3-3). Quantum fuzzy four-sphere $ S_{qF}^{4} $ is the unital $ *- $ algebra (over $\mathbb{C}$) generated by $ (X_{\pm1}^{q}, X_{\pm2}^{q}, X_{0}^{q}) $ satisfying the following relations:
\begin{equation}
X_{+2}^{q}X_{+1}^{q}= qX_{+1}^{q}X_{+2}^{q},\quad
X_{-2}^{q}X_{+1}^{q}= q^{-1}X_{+1}^{q}X_{-2}^{q}, \quad X_{0}^{q*}= X_{0}^{q}\tag{3-15}
\end{equation}
together with the quantum four-sphere constraint 
\begin{equation*}
\mathbf{X}^{q}\cdot \mathbf{X}^{q}= \sum_{n=0,\pm1,\pm2}q^{n}X_{-n}^{q}X_{+n}^{q}= 
\end{equation*}
\begin{equation}
X_{0}^{2}+ qX_{-1}^{q}X_{+1}^{q}+  q^{-1}X_{+1}^{q}X_{-1}^{q}+ q^{2}X_{-2}^{q}X_{+2}^{q}+ q^{-2}X_{+2}^{q}X_{-2}^{q}=1.\tag{3-16}
\end{equation}
In the quantum fuzzy sphere $ S_{qF}^{4} $ we have two different parameter $ q $ and $ l $, which we take to be real. Then, there are different limit cases. These different limits can be expressed as:
\begin{equation*}
 \bullet  first\quad limit:  S_{qF}^{4} \longrightarrow  S_{F}^{4} \longrightarrow S^{4}   ,\;( q\longrightarrow 1  followed\quad by \quad l\longrightarrow \infty )
\end{equation*}
\begin{equation*}
 \bullet  second\quad limit:  S_{qF}^{4} \longrightarrow  S_{q}^{4} \longrightarrow S^{4}   ,\;(l \longrightarrow \infty  followed\quad by \quad q\longrightarrow 1 )
\end{equation*}
\begin{equation}
 \bullet  third \quad limit:  S_{qF}^{4} \longrightarrow  S^{4}   ,\;(l \longrightarrow \infty ,\:  q\longrightarrow 1 \quad simultaneously).\tag{3-17}
\end{equation}

The equation(3-16) in the third limit case reduces to (3-5).

\section{The Hopf fibration on $ S_{qF}^{4} $}
The classical Hopf fibration on $ S^{4} $ is   $SU(2)$ principal fibration  $\pi$ with $S^{7}$ as total manifold over the base manifold $S^{4}$:
\begin{equation}
SU(2) \;\xrightarrow{right SU(2)-action}\; S^{7}\;\xrightarrow{\pi}\; S^{4},
\tag{4-1}
\end{equation}

Let $\mathcal{B}_{\mathbb{C}} =C^{\infty}(S^{7},\mathbb{C})$ and $\mathcal{A}_{\mathbb{C}}= C^{\infty}(S^{4},\mathbb{C})$ denote the algebras of $\mathbb{C}$-valued smooth functions on the total manifold $S^{7}$ and base manifold $ S^{4}$ under point-wise multiplication, respectively. The finite dimensional irreducible representations of the Lie group $ SU(2) $ are classified by a positive integer $ l $ on the carrier space $ \mathcal{V}^{(l)}\simeq Sym^{l}(\mathbb{C}^{2}) $. The elements of $\mathcal{B}_{\mathbb{C}} $ can be classified into the right modules,
\begin{equation}
C_l^{\infty}(S^{7},\mathcal{V}^{(l)}) =\{\varphi :S^{7} \rightarrow\mathcal{V}^{(l)},\quad \varphi (p\cdot\omega)= \omega^{-1} \cdot \varphi(p)\:, \quad \forall p \in S^{7}\: ,\:\forall\omega\in SU(2)\}
\tag{4-2},
\end{equation}

over the pull back of the $\mathcal{A}_{\mathbb{C}}$. The Serre-Swan theorem [27] states that for a compact smooth manifold $ S^{4} $, there is a complete  equivalence between the category of vector bundles over that manifold and bundle maps, and the category of finitely generated projective modules over the algebra $ C(S^{4}) $ of functions over $ S^{4} $ and module morphisms. In algebraic $ K $-theory, it is well known that corresponds to these bundles, there are  projectors $ \mathcal{P}_{l}\in \mathcal{M}_{4^{l}}(C(S^{4}))$ such that, for the associated vector bundle 
\begin{equation}
\mathcal{E}^{(l)} = S^{7} \times_{SU(2)} \mathcal{V}^{(l)} \xrightarrow{\pi} S^{4},
\tag{4-3}
\end{equation}
right $ \mathcal{A}_{\mathbb{C}} $-module of sections $\Gamma^{\infty}(S^{4},\mathcal{E}^{(l)})$ which is isomorphic with  $C_{(l)}^{\infty}(S^{7}, \mathcal{V}^{(l)})$ is equivalent to the image in the free module $ (\mathcal{A}_{\mathbb{C}})^{4^{l}}$ of a projector $ \mathcal{P}_{l} $, $\Gamma^{\infty}(S^{4},\mathcal{E}^{(l)})=\mathcal{P}_{l}(\mathcal{A}_{\mathbb{C}})^{4^{l}} $. The projector $ \mathcal{P}_{l} $ is a Hermitian operator of rank $1$. These projectors are $ 4^{l}\times 4^{l} $ matrices taking values in $ C^{\infty}(S^{4}) $ such that $ \Gamma^{\infty}(S^{4}, \mathcal{E}^{l})= \mathcal{P}_{l}(C^{\infty}(S^{4}))^{4^{l}} $ which is isomorphic with $C_l^{\infty}(S^{7},\mathcal{V}^{(l)})$ as right $ C^{\infty}(S^{4})- $module:
\begin{equation}
\mathcal{P}_{l}\in Mat_{4^{l}} (\mathcal{A}_{\mathbb{C}}),\quad \mathcal{P}_{l}^{2}=\mathcal{P}_{l},\quad \mathcal{P}_{l}^{\dagger} =  \mathcal{P}_{l},\quad Tr \mathcal{P}_{l}=1.
\tag{4-4}
\end{equation} 
From now on, we focus on the case $ l=1 $ with $ \mathcal{V}^{(1)}= \mathbb{C}^{2} $.
For the right $ \mathcal{A}_{\mathbb{C}} $-module of sections  $ \Gamma^{\infty}(S^{4},\mathcal{E}^{(\pm)}) $ there exist two projectors $ \mathcal{P}_{+}$ and $\mathcal{P}_{-} $ having the same rank $ 1 $. Therefore, the free module $ (\mathcal{A}_{\mathbb{C}})^{4} $ can be written as a direct sum of the projective $ \mathcal{A}_{\mathbb{C}}- $modules,
\begin{equation}
(\mathcal{A}_{\mathbb{C}})^{4}= \mathcal{P}_{+}(\mathcal{A}_{\mathbb{C}})^{4}\bigoplus \mathcal{P}_{-}(\mathcal{A}_{\mathbb{C}})^{4},
\tag{4-5}
\end{equation}
where $ \mathcal{P}_{+}+\mathcal{P}_{-}=1 $.\\
Now, let us briefly describe the quantum version of the Hopf fibration $ S^{7}\rightarrow S^{4} $. The quantum fuzzy seven-sphere $ S_{qF}^{7} $ is the total manifold of a quantum $ SU_{q}(2) $ instanton bundle over a quantum fuzzy four-sphere $ S_{qF}^{4} $: $ S_{qF}^{7}\rightarrow S_{qF}^{4} $. We denote the coordinate algebra of $ S_{qF}^{7}, S_{qF}^{4} $ and $ SU_{q}(2) $ by $ \mathcal{A}( S_{qF}^{7}), \mathcal{A}( S_{qF}^{4}) $ and $ \mathcal{A}(SU_{q}(2)) $, respectively. $ \mathcal{A}(SU_{q}(2)) $ is a Hopf algebra. Also, $ \mathcal{A}( S_{qF}^{7}) $ is a right $ \mathcal{A}(SU_{q}(2))- $comodule algebra with multiplication $ m: \mathcal{A}( S_{qF}^{7})\otimes \mathcal{A}( S_{qF}^{7})\rightarrow \mathcal{A}( S_{qF}^{7}) $, and coaction $ \Delta_{R}: \mathcal{A}( S_{qF}^{7})\rightarrow \mathcal{A}( S_{qF}^{7})\otimes \mathcal{A}(SU_{q}(2)) $. It has been showed that [38] the extension $ \mathcal{A}( S_{qF}^{4})\hookrightarrow \mathcal{A}( S_{qF}^{7}) $ is a $ \mathcal{A}(SU_{q}(2))- $ Hopf-Galois extension, which is not cleft(is not trivial). The Hopf-Galois extension $ \mathcal{A}( S_{qF}^{4})\hookrightarrow \mathcal{A}( S_{qF}^{7}) $ is the quantum version of the second Hopf fibration $ S^{7}\rightarrow S^{4} $. For the Hopf-Galois extension $ \mathcal{A}( S_{qF}^{4})\hookrightarrow \mathcal{A}( S_{qF}^{7}) $, let $ \rho: \mathbb{C}^{2}\rightarrow  \mathbb{C}^{2}\otimes \mathcal{A}(SU_{q}(2)) $ be the fundamental corepresentation of $ \mathcal{A}(SU_{q}(2)) $. Also, let $ C( \mathcal{A}( S_{qF}^{7}),\mathbb{C}^{2} ) $ be the right $ \mathcal{A}( S_{qF}^{4})- $module of coequivariant maps. The quantum projectors $ \mathcal{P}_{\pm}\in Mat_{4}( \mathcal{A}( S_{qF}^{4}))\simeq Mat_{4}(\mathbb{C})\otimes \mathcal{A}(S_{qF}^{4}) $ determine  quantum fuzzy vector bundles over $ S_{qF}^{4} $, whose module of sections are $ \mathcal{P}_{\pm}(\mathcal{A}(S_{qF}^{4}))^{4}\simeq \Gamma_{\rho}(\mathcal{A}( S_{qF}^{4}),\mathbb{C}^{2} )$. Let us define $ (\mathcal{A}(S_{qF}^{4}))^{4}:= \mathcal{A}(S_{qF}^{4})\otimes \mathbb{C}^{4} $. It is clear that $ \mathcal{P}_{\pm}(\mathcal{A}(S_{qF}^{4}))^{4} $ are q-deformation of the classical instanton bundles over commutative four-sphere $ S^{4} $. It means that in the limit case when the parameters $ q $ and $ l $ tend to $ 1 $ and infinity, respectively, the quantum projective modules $  \mathcal{P}_{\pm}(\mathcal{A}(S_{qF}^{4}))^{4}  $ tend to $  \mathcal{P}_{\pm}(\mathcal{A}(S^{4}))^{4}  $, the modules of sections of the complex rank two instanton bundles on commutative four-sphere. Also, it has been showed that $  \mathcal{P}_{+}(\mathcal{A}(S_{qF}^{4}))^{4}\simeq \mathcal{P}_{-}(\mathcal{A}(S_{qF}^{4}))^{4}\simeq \bigoplus_{l\in \mathbb{N}+1/2}\mathcal{V}^{(l)} $ as $ \mathcal{U}_{q}(so(5)) $ representations[38].

\section{Spin $\frac{1}{2}$ q-deformed projectors of the left and right q-deformed projective $ A(S_{qF}^{4}) -$module}

As mentioned before, according to Serre-Swan theorem [27], for a compact smooth manifold $ M $, there is a complete  equivalence between the category of vector bundles over that manifold and bundle maps, and the category of finitely generated projective modules over the algebra $ C(M) $ of functions over $ M $ and module morphisms. Therefore, study of the quantum principal fibration $S_{qF}^{7}\xrightarrow{SU_{q}(2)} S_{qF}^{4}$, replaces with the study of noncommutative finitely generated projective $ A(S_{qF}^{4})  -$module of its sections. To build the left and right q-deformed projective modules we should construct the fuzzy q-projectors of these modules.
As we mentioned before, $ \sigma $ in () is unitary spin $ (1/2,1/2) *-$representation of $ \mathcal{U}_{q}(so(5)) $ on the vector space $ \mathcal{V}^{(1/2)} $.  Now, using the decomposition $ \mathcal{V}^{(l)}\otimes     \mathcal{V}^{(1/2)}\simeq \mathcal{V}^{(l+ 1/2)}\oplus \mathcal{V}^{(l- 1/2)} $, the projectors for left projective module can be written as:
\begin{equation}
\mathcal{P}_{[l\pm\frac{1}{2}]_{q}}^{L} = \frac{1}{[2]_{q}} [1 \pm \dfrac{\mathbf{\Sigma}\cdot \mathbf{J}^{L}+1}{\sqrt{[l]_{q}[l+3]_{q}}•}] 
\tag{5-1}
\end{equation}
where $ \mathbf{\Sigma} $ and $ \mathbf{J} $ are spin $ 1/2 $ and $ L/2 $ representations of $ \mathcal{U}_{q}(so(5)) $, respectively and $ \mathbf{\Sigma}\cdot \mathbf{J} $ is as follow:
\begin{equation*}
\mathbf{\Sigma}\cdot \mathbf{J}= \sum_{n=0,\pm1,\pm2}q^{n}\Sigma_{-n}J_{+n}= 
\end{equation*}
\begin{equation}
\Sigma_{0}J_{0}+ q\Sigma_{-1}J_{+1}+  q^{-1}\Sigma_{+1}J_{-1}+ q^{2}\Sigma_{-2}J_{+2}+ q^{-2}\Sigma_{+2}J_{-2}
\tag{5-2}
\end{equation}
which couples left angular momentum and spin $ \dfrac{1}{2} $ to its maximum and minimum values $ \l \pm \dfrac{1}{2}$. Here, $ \mathbf{\Sigma}_{i} $ are spin $ 1/2 $ q-deformed $ \mathcal{U}_{q}(so(5)) $ generators and $ \mathbf{J}_{i}^{L} $ are left q-deformed $ l/2- $representation of $ \mathcal{U}_{q}(so(5)) $. 

It is easy to see that
\begin{equation}
\mathcal{P}_{[l+ \frac{1}{2}]_{q}}^{L} + \mathcal{P}_{[l-\frac{1}{2}]_{q}}^{L} = \frac{2}{[2]_{q}}. 
\tag{5-3}
\end{equation}
These are the projectors of our left projective $ \mathcal{A}(S_{qF}^{4})  -$module 
\begin{equation}
(\mathcal{A}(S_{qF}^{4}))^{4}=(\mathcal{A}(S_{qF}^{4}))^{4}\mathcal{P}_{[l+ \frac{1}{2}]_{q}}^{L}\oplus (\mathcal{A}(S_{qF}^{4}))^{4}\mathcal{P}_{[l- \frac{1}{2}]_{q}}^{L}.\tag{5-4}
\end{equation}
Equation (5-3), in the limit case $ q\longrightarrow 1 $ reduces to the following condition:
\begin{equation}
\mathcal{P}_{l+ \frac{1}{2}}^{L} + \mathcal{P}_{l-\frac{1}{2}}^{L} = 1.
\tag{5-5}
\end{equation}
Using (5-1) we can define the corresponding q-deformed idempotents as follow:
\begin{equation}
\Gamma_{[l\pm \frac{1}{2}]_{q}}^{L}=[2]_{q}\mathcal{P}_{[l\pm \frac{1}{2}]_{q}}^{L}-1=\pm \frac{\mathbf{\Sigma} \cdot \mathbf{J}^{L}+1}{\sqrt{•[l]_{q}[l+3]_{q}}}
\tag{5-6}
\end{equation}

In the limit $ q\longrightarrow 1 $ (5-1) and (5-6) become:
\begin{equation}
\mathcal{P}_{(l\pm\frac{1}{2})}^{L} = \frac{1}{2} [1 \pm \dfrac{\mathbf{G}\cdot \mathbf{G}^{(l)L}+1}{\sqrt{l(l+3)}•}] 
\tag{5-7}
\end{equation}
\begin{equation}
\Gamma_{l\pm \frac{1}{2}}^{L}=\pm \dfrac{\mathbf{G}\cdot \mathbf{G}^{(l)L}+1}{\sqrt{l(l+3)}•}
\tag{5-8}
\end{equation}
where $  \mathbf{G}\cdot \mathbf{G}^{(l)L}= G_{i}G_{i}^{(l)L}$. The projector $\mathcal{P}_{[l\pm \frac{1}{2}]_{q}}^{R} $ coupling the right generator and spin $ \dfrac{1}{2} $ to its maximum and minimum values $ l\pm \dfrac{1}{2} $ is obtained by changing $ \mathbf{J}^{L} $ to $ -\mathbf{J}^{R} $ in the above expression
\begin{equation}
\mathcal{P}_{[l\pm\frac{1}{2}]_{q}}^{R} = \frac{1}{[2]_{q}} [1 \pm \dfrac{-\mathbf{\Sigma}\cdot \mathbf{J}^{R}+1}{\sqrt{[l]_{q}[l+3]_{q}}•}] 
\tag{5-9}
\end{equation}

Here, again it is easy to see that
\begin{equation}
\mathcal{P}_{[l+ \frac{1}{2}]_{q}}^{R} + \mathcal{P}_{[l-\frac{1}{2}]_{q}}^{R} = \frac{2}{[2]_{q}}. 
\tag{5-10}
\end{equation}
These are the projectors of our right $ \mathcal{A}(S_{qF}^{4})  -$module 
\begin{equation}
(\mathcal{A}(S_{qF}^{4}))^{4}=\mathcal{P}_{[l+ \frac{1}{2}]_{q}}^{R} (\mathcal{A}(S_{qF}^{4}))^{4}\oplus \mathcal{P}_{[l- \frac{1}{2}]_{q}}^{R} (\mathcal{A}(S_{qF}^{4}))^{4}.\tag{5-11}
\end{equation}
The corresponding quantum idempotents are
\begin{equation}
\Gamma_{[l\pm \frac{1}{2}]_{q}}^{R}=[2]_{q}\mathcal{P}_{[l\pm \frac{1}{2}]_{q}}^{R}-1=\pm \frac{-\mathbf{\Sigma} \cdot \mathbf{J}^{R}+1}{\sqrt{•[l]_{q}[l+3]_{q}}}
\tag{5-12}
\end{equation}
It is clear that
\begin{equation}
\lim_{q \to 1 ,\; l \to \infty} (\Gamma_{[l\pm\frac{1}{2}]_{q}}^{L,R})^{2} = 1.
\tag{5-13}
\end{equation}
In the limit $ q\longrightarrow 1 $, (5-9) and (5-12) become:
\begin{equation}
\mathcal{P}_{(l\pm\frac{1}{2})}^{R} = \frac{1}{2} [1 \pm \dfrac{-\mathbf{G}\cdot \mathbf{G}^{(l)R}+1}{\sqrt{l(l+3)}•}] 
\tag{5-14}
\end{equation}
\begin{equation}
\Gamma_{l\pm \frac{1}{2}}^{R}=\pm \dfrac{-\mathbf{G}\cdot \mathbf{G}^{(l)R}+1}{\sqrt{l(l+3)}•}
\tag{5-15}
\end{equation}
which are the results of fuzzy sphere limit.

\section{q-deformed fuzzy Ginsparg-Wilson algebra and its Spin $\frac{1}{2}$ q-deformed fuzzy Dirac and chirality operators}
 The q-deformed fuzzy Ginsparg-Wilson algebra $ \mathcal{A}_{qF}  $ is the $ \ast $ -algebra over $ \mathbb{C} $, generated by two $ \ast $ -invariant q-deformed involution $ \Gamma^{qF} $ and ${\Gamma^{\prime}}^{qF}  $:
\begin{equation}
\mathcal{A}_{qF} = \langle\Gamma^{qF},{\Gamma^{\prime}}^{qF}\colon\quad(\Gamma^{qF})^{2} =({\Gamma^{\prime}}^{qF})^{2}=\textit{I},\quad (\Gamma^{qF})^{\dagger}=\Gamma^{qF},\quad ({\Gamma^{\prime}}^{qF})^{\dagger}={\Gamma^{\prime}}^{qF}\rangle , 
\tag{6-1}
\end{equation}
each representation of (6-1) is a realization of the q-deformed Ginsparg-Wilson algebra.
Now, consider the following two elements constructed out of the generators $ \Gamma^{qF} $ and ${\Gamma^{\prime}}^{qF}  $ of the q-deformed fuzzy Ginsparg-Wilson algebra $ \mathcal{A}_{qF}  $:
\begin{equation}
\begin{split}
\Gamma_{1}^{qF} = \Gamma^{qF} + {\Gamma^{\prime}}^{qF} \; , \qquad \qquad {(\Gamma_{1}^{qF}})^{*} = \Gamma_{1}^{qF} ,\\
\Gamma_{2}^{qF} = \Gamma^{qF} - {\Gamma^{\prime}}^{qF} \; , \qquad \qquad {(\Gamma_{2}^{qF}})^{*} =  \Gamma_{2}^{qF} .
\end{split}
\tag{6-2}
\end{equation}
So that, $\Gamma_{1}^{qF}$ and $\Gamma_{2}^{qF}$ anticommute with each other:
\begin{equation}
\left\lbrace \Gamma_{1}^{qF} , \Gamma_{2}^{qF}\right\rbrace  = 0.
\tag{6-3}
\end{equation}
Identifying $ \Gamma_{[l\pm \frac{1}{2}]_{q}} ^{L}$ and $ \Gamma_{[l\pm \frac{1}{2}]_{q}} ^{R}$ with $ \Gamma^{qF} $ and ${\Gamma^{\prime}}^{qF}  $, we get:
\begin{equation}
\Gamma_{1}^{qF\pm }=\pm \frac{\mathbf{\Sigma} \cdot \mathbf{\mathcal{J}}+2}{\sqrt{•[l]_{q}[l+3]_{q}}}, \qquad 
\Gamma_{2}^{qF\pm }= \pm \frac{\mathbf{\Sigma} \cdot (\mathbf{J}^{L}+\mathbf{J}^{R})}{\sqrt{•[l]_{q}[l+3]_{q}}}.
\tag{6-4}
\end{equation}
where we defined $\mathbf{\mathcal{J}}:=\mathbf{J}^{L}-\mathbf{J}^{R}  $ is the q-deformed fuzzy version of angular momentum operator.

Now, let us define the q-deformed fuzzy Dirac and chirality operators on q-deformed fuzzy four-sphere $ S_{qF}^{4} $ as follow:
\begin{equation}
\mathcal{D}_{qF}^{\pm}=\sqrt{•[l]_{q}[l+3]_{q}}\Gamma_{1}^{qF\pm }=\pm (\mathbf{\Sigma} \cdot \mathcal{J}+ 2),\qquad 
\gamma _{qF}^{\pm}=\dfrac{1}{2•}\Gamma_{2}^{qF\pm }
\tag{6-5}
\end{equation}
Here, let us apply the third limit case mentioned in(3-17) to (6-5):
\begin{equation}
 \lim_{q \to 1,l \to \infty} \mathcal{D}_{qF}^{\pm}=\pm(\mathbf{G} \cdot \mathbf{\mathcal{G}}^{(l)}+2),\qquad  \lim_{q \to 1,l \to \infty}\gamma_{qF}^{\pm} =\pm \mathbf{G} \cdot \mathcal{\mathbf{x}}.
 \tag{6-6}
\end{equation}
These are correct Dirac and chirality operators on commutative $ S^{4} $. Also, it is easy to see that:
\begin{equation}
 \lim_{q \to 1,l\to \infty} \lbrace \mathcal{D}_{qF}^{\pm},\gamma_{qF}^{\pm}\rbrace=0,
 \tag{6-7}
\end{equation}
which we expect from Dirac and chirality operators on $ S^{4} $.

These $\mathcal{D}_{qF}^{\pm}  $ and $\gamma_{qF}^{\pm}  $ are not the only Dirac and chirality operators on $ S_{qF}^{4} $. In noncommutative geometry, the right and left $ \mathcal{A}(S_{qF}^{4}) -$modules are not isomorphic. 
Because the left and right momentum operators are not equivalent, we can choose another set of operators to construct our q-deformed Ginsparg-Wilson algebra.

Choosing $ \Gamma_{[l+\frac{1}{2}]_{q}}^{R} $ and $ \Gamma_{[l-\frac{1}{2}]_{q}}^{L} $ and considering $ \sqrt{[l]_{q}[l+3]_{q}}(\Gamma_{[l+\frac{1}{2}]_{q}}^{R}-\Gamma_{[l-\frac{1}{2}]_{q}}^{L}) $, we get the correct Dirac operator (6-6) in the limit case. The corresponding chirality operator is got from  $ (\Gamma_{[l+\frac{1}{2}]_{q}}^{R}+\Gamma_{[l-\frac{1}{2}]_{q}}^{L}) $ as this goes to the correct limit case (6-6). The other possibility is to combine $ \Gamma_{[l-\frac{1}{2}]_{q}}^{R}  $ and $ \Gamma_{[l+\frac{1}{2}]_{q}}^{L} $, we define  $ -\sqrt{[l]_{q}[l+3]_{q}}(\Gamma_{[l-\frac{1}{2}]_{q}}^{R}-\Gamma_{[l+\frac{1}{2}]_{q}}^{L}) $ and $ (\Gamma_{[l-\frac{1}{2}]_{q}}^{R}+\Gamma_{[l+\frac{1}{2}]_{q}}^{L}) $ as Dirac and chirality operators, respectively. These are all the possible combinations.

\section{q-deformed fuzzy gauged Dirac operator ( no instanton fields)}

Let us denote by $ A^{L} $ the  connection $ 1- $form associated with the projector $ \mathcal{P} $ on $ S_{F}^{4} $,
\begin{equation}
 A^{L}\in End_{\mathbb{C}}( C^{\infty}(S^{7}),\mathbb{C})\otimes_{\mathbb{C}}\Omega^{1}(( S^{7}),\mathbb{C})
 \tag{7-1}
\end{equation}
The canonical connection associated with the projector $ \mathcal{P} $ is as follow:
\begin{equation}
\nabla= \mathcal{P}od: \Gamma^{\infty}(S_{qF}^{4},\mathcal{V}^{(l)})\rightarrow \Gamma^{\infty}(S_{qF}^{4},\mathcal{V}^{(l)})\otimes_{C^{\infty}(S_{qF}^{4},\mathbb{C})}\Omega^{1}(S_{qF}^{4},\mathbb{C}),\tag{7-2}
\end{equation}
and the corresponding curvature $ 2- $form is given as the following relation:
\begin{equation}
F= \nabla^{2}: \Gamma^{\infty}(S_{qF}^{4},\mathcal{V}^{(l)})\rightarrow \Gamma^{\infty}(S_{qF}^{4},\mathcal{V}^{(l)})\otimes _{C^{\infty}(S_{qF}^{4},\mathbb{C})}\Omega^{2}(S_{qF}^{4},\mathbb{C}).\tag{7-3}
\end{equation}
Let us start with $ S^{4}_{qF}  (d_{l})\otimes \mathbb{C}^{k} $. The $ *- $invariant fuzzy gauge field $ A_{i}^{L} $ acts on $ \xi = ( \xi_{1} ,.....,\xi_{k}),\xi_{i} \in  S^{4}_{qF}  (d_{l}) $ as:
\begin{equation}
(A_{i}^{L}\xi)_{m}=(A_{i})_{mn}\xi_{n}.
\tag{7-4}
\end{equation}
The $ * $-invariant condition on $ A_{i}^{L} $ is:
\begin{equation}
(A_{i}^{L})^{*} = A_{i}^{L},
\tag{7-5}
\end{equation}
which on the commutative $ S^{4} $ becomes a commutative field $ \mathbf{a}$ and its components $ a_{i} $ have to be tangent to commutative $ S^{4} $: 
\begin{equation}
\mathbf{x}\cdot \mathbf{a} = 0.
\tag{7-6}
\end{equation}
We need a condition to get the above result for large $ l $. One of the conditions of such a nature is:
\begin{equation*}
(\mathbf{J}^{L} + \mathbf{A}^{L}) \cdot (\mathbf{J}^{L} + \mathbf{A}^{L}) = 
\end{equation*}
\begin{equation}
\sum_{m=0,\pm1,\pm2} q^{m}(J^{L} + A^{L})_{-m} (J^{L} + A^{L})_{m} = \mathbf{J}^{L} \cdot \mathbf{J}^{L} = [l]_{q}[l+3]_{q}.
\tag{7-7}
\end{equation}
In the following relations we use the Einstein's sumation convention.
The expansion of (7-7) is:
\begin{equation}
q^{m} J_{-m}A_{m}+q^{m}A_{-m}J_{m}+q^{m}A_{-m}A_{m}=0.
\tag{7-8}
\end{equation}
In the limit case (7-7) reduces to:
\begin{equation}
\lim_{q \to 1}(\mathbf{J}^{L} + \mathbf{A}^{L}) \cdot (\mathbf{J}^{L} + \mathbf{A}^{L})=l(l+3).
\tag{7-9}
\end{equation}
When the parameter $ l $ tends to infinity, $ \dfrac{A_{i}^{L}}{l} $ tends to zero. For large $ l $, the (7-8) gives:
\begin{equation}
q^{m}x_{-m}^{L} A_{m}^{L} +q^{m} A_{-m}^{L} x_{m}^{L} +q^{m} \dfrac{A_{-m}^{L} A_{m}^{L}}{l} = 0.
\tag{7-10}
\end{equation}
$ A_{i}^{L}$ is to remain bounded as $ l $ tends to infinity. Also, in this limit $ x_{i}^{L} $ tends to $ \hat{x}_{i}$, the unit normal to the $ S^{2} $ at $ \hat{\mathbf{x}} $. So in the limiting case, if $ A_{i}^{L} $ tends to $ a_{i} $ then $ \hat{\mathbf{x}} \cdot \mathbf{a}=0 $.\\
Now, we can introduce the q-deformed gauged Ginsparg-Wilson system as follow: We can set:
\begin{equation}
\Gamma^{qF}(\mathbf{A}^{L}) = \dfrac{\mathbf{\Sigma}\cdot (\mathbf{J}^{L} + \mathbf{A}^{L})+1}{|\mathbf{\Sigma}\cdot (\mathbf{J}^{L} + \mathbf{A}^{L})+1|}.
\tag{7-11}
\end{equation}
It is an involutory and $ * $-invariant operator:
\begin{equation}
\Gamma^{qF}(\mathbf{A}^{L})^{2} = 1 ,\qquad \Gamma^{qF}(\mathbf{A}^{L})^{*} = \Gamma^{qF}(\mathbf{A}^{L}).
\tag{7-12}
\end{equation}
The gauged involution (7-11), reduces to (5-6) for zero $ \mathbf{A}^{L}$. We put $ \Gamma^{qF} = \Gamma^{qF}(\mathbf{A}^{L}=0) $.\\ Also, we can define the second gauged involution as:
\begin{equation}
\Gamma^{'qF}(\mathbf{A}^{L}) = \Gamma^{'qF}(0)=\Gamma^{'qF}.
\tag{7-13}
\end{equation}
We put $ \Gamma^{'}=\Gamma^{'}(A^{L}=0) $. Notice that, the operators $ \mathbf{J}^{L,R} $  do not have continuum limit as their squares $ l(l+3)$ diverege as $ l $ tends to infinity. In contrast, $\boldsymbol{\mathcal{J}} $ and $ \mathbf{A}^{L} $ do have continuum limits.\\
 Up to the first order, (7-11) can be written as follow:
\begin{equation}
\Gamma^{qF}(\mathbf{A}^{L}) = \dfrac{ \mathbf{\Sigma}\cdot (\mathbf{J}^{L} + \mathbf{A}^{L})+1}{\sqrt{[l]_{q}[l+3]_{q}}}.
\tag{7-14}
\end{equation}
and
\begin{equation}
\Gamma^{'qF} = \dfrac{\mathbf{\Sigma}\cdot \mathbf{J}^{R} +1 }{\sqrt{[l]_{q}[l+3]_{q}}}.
\tag{7-15}
\end{equation}
In the limit case when the parameter q tends to unit, (7-14) tends to:
\begin{equation}
 \lim_{q \to 1}\Gamma^{qF}(\mathbf{A}^{L})=\frac{\mathbf{G} \cdot (\mathbf{G}^{(l)L}+ \mathbf{A}^{L})+1}{\sqrt{l(l+3)}}
\tag{7-16}
\end{equation}
Now, we can construct the following $ *- $invariant operators:
\begin{equation}
\begin{split}
\Gamma_{1}^{qF}(\mathbf{A}^{L}) =\Gamma^{qF}(\mathbf{A}^{L}) + {\Gamma^{\prime}}^{qF} \; , \qquad \qquad {(\Gamma_{1}^{qF}})^{*} = \Gamma_{1}^{qF} ,\\
\Gamma_{2}^{qF}(\mathbf{A}^{L}) =\Gamma^{qF}(\mathbf{A}^{L}) - {\Gamma^{\prime}}^{qF} \; , \qquad \qquad {(\Gamma_{2}^{qF}})^{*} =  \Gamma_{2}^{qF} .
\end{split}
\tag{7-17}
\end{equation}
Let us define the gauged q-deformed fuzzy Dirac and chirality operators on q-deformed fuzzy four-sphere $ S_{qF}^{4} $ as follow:
\begin{equation}
D_{qF}^{\pm}(\mathbf{A}^{L})=\sqrt{[l]_{q}[l+3]_{q}}\Gamma_{1}^{qF}(\mathbf{A}^{L})=\pm (\mathbf{\Sigma} \cdot (\mathbf{•\mathcal{J}}+\mathbf{A}^{L})+ 2),\qquad 
\tag{7-18}
\end{equation}
and for chirality operator:
\begin{equation}
\gamma _{qF}(\mathbf{A}^{L})=\Gamma_{2}^{qF}(\mathbf{A}^{L})= \pm \dfrac{ \mathbf{\Sigma}\cdot (\mathbf{J}^{L}+\mathbf{ J}^{R} + \mathbf{A}^{L})}{\sqrt{[l]_{q}[l+3]_{q}}}.
\tag{7-19}
\end{equation}
Here, let us apply the third limit case mentioned in (3-17) to (7-18) and (7-19):
\begin{equation}
 \lim_{q \to 1,l \to \infty} D_{qF}^{\pm}(\mathbf{A}^{L})=\pm(\mathbf{G} \cdot (\mathbf{\mathcal{G}}^{(l)}+\mathbf{A}^{L})+2),\qquad  \lim_{q \to 1,l \to \infty}\gamma_{qF}(\mathbf{A}^{L}) =\pm \mathbf{G} \cdot \mathcal{\mathbf{x}}.
 \tag{7-20}
\end{equation}
These are the correct gauged Dirac and chirality operators on commutative $ S^{4} $

\section{Instanton coupling}
According to the Serre-Swan theorem, study of the quantum principal fibration $S_{qF}^{7}\;\xrightarrow{SU_{q}(2)}\; S_{qF}^{4} $, are replaced with the study of noncommutative finitely generated projective module of the Hopf-Galois extension $ \mathcal{A}( S_{qF}^{4})\hookrightarrow  \mathcal{A}(S_{qF}^{7})$. To build the projective module, let introduce  $\mathbb{C}^{d_{t}}$ with $ d_{t}= \dfrac{1}{6•}(t+1)(t+2)(t+3) $ carrying the $ (t/2,t/2) *- $representation of angular momentum of $ \mathcal{U}_{_{q}}(so(5))$. Here, the algebra $ \mathcal{U}_{q}(so(5))\equiv so_{q}(5) $ is generated by elements $ T_{\pm1},T_{\pm2} $ and $ T_{\pm0} $ satisfying the following relations:
\begin{equation*}
[T_{+1}, T_{-1}]= \dfrac{T_{+0}^{2}- T_{+0}^{-2}}{q-q^{-1}•},\quad
[T_{+2}, T_{-2}]= \dfrac{T_{-0}^{2}- T_{-0}^{-2}}{q^{2}-q^{-2}•}
\end{equation*}
\begin{equation*}
[T_{+1}, T_{-2}]= [T_{+2}, T_{-1}]=0, \quad T_{\pm 1}^{*}= T_{\mp 1}, T_{\pm 2}^{*}= T_{\mp 2},
\end{equation*}
\begin{equation*}
T_{+0}T_{\pm1}= q^{\pm 1}T_{\pm 1}T_{+0},\quad  T_{+0}T_{\pm 2}= q^{\mp 1}T_{\pm 2}J_{+0}
\end{equation*}
\begin{equation*}
T_{-0}T_{\pm2}= q^{\pm 1}T_{\pm 2}T_{-0},\quad  T_{-0}T_{\pm 1}= q^{\mp 1}T_{\pm 1}J_{-0}
\end{equation*}
\begin{equation*}
T_{\pm1}^{3}T_{\pm2}- (q^{2}+q^{-2}+1)T_{\pm1}^{2}T_{\pm2}T_{\pm1}+ (q^{2}+q^{-2}+1)T_{\pm1}T_{\pm2}T_{\pm1}^{2}- T_{\pm2}T_{\pm1}^{3}=0,
\end{equation*}
\begin{equation}
T_{\pm2}^{2}T_{\pm1}- (q^{2}+q^{-2})T_{\pm2}T_{\pm1}T_{\pm2}+ T_{\pm1}T_{\pm2}^{2}=0.\tag{8-1}
\end{equation}
The coproduct $ \Delta: \mathcal{U}_{q}(so(5))\rightarrow \mathcal{U}_{q}(so(5))\otimes \mathcal{U}_{q}(so(5)) $, counit $ \varepsilon: \mathcal{U}_{q}(so(5))\rightarrow \mathbb{C} $ and antipode $ S: \mathcal{U}_{q}(so(5))\rightarrow \mathcal{•U}_{q}(so(5)) $
 of the Hopf algebra $ \mathcal{U}_{q}(so(5)) $ are given by:
\begin{equation*}
\Delta T_{\pm 0}= T_{\pm 0}\otimes T_{\pm 0},
\end{equation*}
\begin{equation*}
\Delta T_{\pm i}= T_{\pm i}\otimes T_{\pm 0}+ T_{\pm 0}\otimes T_{\pm i}, \quad i=1,2,
\end{equation*}
\begin{equation*}
\varepsilon(T_{\pm 0})=1,\quad \varepsilon(T_{\pm i})= 0, \quad i=1,2,
\end{equation*}
\begin{equation}
S(T_{\pm 0})= T_{\pm 0}^{-1},\quad S(T_{\pm i})= -q^{i}T_{\pm i}, \quad i=1,2,\tag{8-2}
\end{equation}
where the $ \ast $ -structure is for real $ q $. 
Also, let, $ \mathcal{P}_{qF}^{[l+t]_{q}} $ be the  projector coupling left angular momentum operator $ \mathbf{J}^{L}$ with $ \mathbf{T} $ to produce maximum angular momentum $ l+t $. We know that the image of a  projector on a free module is a projective module. Then, as $ Mat(d_{l})^{d_{t}}=Mat (d_{l})\otimes \mathbb{C}^{d_{t}} $ is a free module, therefore, $  \mathcal{P}_{qF}^{[l+t]_{q}} Mat(d_{l})^{d_{t}} $ is the fuzzy version of $ SU(2) $ bundle on $S_{qF}^{4}$. Also, we can use the projector $ \mathcal{P}_{qF}^{[l-t]_{q}} $ to produce the projective module $ \mathcal{P}_{qF}^{[l-t]_{q}} Mat (d_{l})^{d_{t}}$ to introduce the least angular momentum $ (l-t) $.\\ The q-deformed fuzzy projectors $\mathcal{P}_{qF}^{[l\pm t]_{q}}$ corresponding to $ (l\pm t)$-representations of $ so_{q}(5) $ can be written as:
\begin{equation}
\mathcal{P}_{[l\pm t]_{q}}^{L} = \frac{1}{[2]_{q}} [1 \pm \dfrac{\mathbf{\Sigma}\cdot (\mathbf{J}^{L}+\mathbf{T})+1}{\sqrt{[l\pm t]_{q}[(l\pm t)+3]_{q}}•}], \quad \mathcal{P}_{_{qF}}^{(l \pm t)^{*}}=P_{qF}^{(l \pm t)},
\tag{8-3}
\end{equation}
\begin{equation}
Mat(d_{l})\otimes\mathbb{C}^{d_{t}}=(Mat(d_{l})\otimes\mathbb{C}^{d_{t}})\mathcal{P}_{qF}^{(l+t)}\oplus  (Mat(d_{l})\otimes \mathbb{C}^{d_{t}})\mathcal{P}_{qF}^{(l-t)}.
\tag{8-4}
\end{equation}
To set the q-deformed fuzzy Ginsparg-Wilson system in instanton sector, we choose the following $ *-$invariant involution $ \Gamma $ for the highest and lowest weights $ l\pm t $:
\begin{equation*}
\Gamma_{qF}^{\pm}(\mathbf{T})=[2]_{q}\mathcal{P}_{qF}^{(l\pm t)}-1=\pm \dfrac{\mathbf{\Sigma}\cdot(\mathbf{J}^{L}+\mathbf{T})+1}{\sqrt{[l\pm t]_{q}[(l\pm t)+3]_{q}}}
\end{equation*}
\begin{equation}
(\Gamma_{qF}^{\pm}(\mathbf{T}))^{2}=1,\quad \Gamma_{qF}^{\pm^{*}}(\mathbf{T})=\Gamma_{_{qF}}^{\pm}(\mathbf{T}).
\tag{8-5}
\end{equation}
It is clear that $ \Gamma_{qF}^{\pm}(T=0)=\Gamma_{qF} $. On the module $( Mat(d_{l})^{d_{t}}\otimes\mathbb{C}^{2})\mathcal{P}_{qF}^{(l \pm t)} $ we have:
\begin{equation}
(\mathbf{J}^{L}+\mathbf{T})^{2} =[(l \pm t)]_{q}[(l \pm t)+3]_{q}.
\tag{8-6}
\end{equation}
The quantum projectors $\mathcal{P}_{[l\pm t]_{q}}^{R} $ coupling the right momentum and instanton to its maximum and minimum values $ (l\pm t)\pm \dfrac{1}{2} $, respectively are obtained by changing $ J_{i}^{L} $ to $ -J_{i}^{R} $ in the above expression
 \begin{equation}
\mathcal{P}_{[l\pm t]_{q}}^{R} = \frac{1}{[2]_{q}} [1 \pm \dfrac{\mathbf{\Sigma}\cdot (-\mathbf{J}^{R}+\mathbf{T})+1}{\sqrt{[l\pm t]_{q}[(l\pm t)+3]_{q}}•}], \quad \mathcal{P}_{_{qF}}^{(l \pm t)^{*}}=P_{qF}^{(l \pm t)},
\tag{8-7}
\end{equation}
 These are the quantum projectors of our right projective $ \mathcal{A}(S_{qF}^{4})  -$module 
\begin{equation}
(\mathcal{A}(S_{qF}^{4}))^{4}=(\mathcal{A}(S_{qF}^{4}))^{4}\mathcal{P}_{[(l\pm t)+ \frac{1}{2}]_{q}}^{R}\bigoplus (\mathcal{A}(S_{qF}^{4}))^{4}\mathcal{P}_{[(l\pm t)- \frac{1}{2}]_{q}}^{R}.\tag{8-8}
\end{equation}
 The corresponding quantum idempotents are
 \begin{equation}
\Gamma_{qF}^{\pm R}(\mathbf{T})=[2]_{q} \mathcal{P}_{[l\pm t]_{q}}^{R}-1=\pm \dfrac{\mathbf{\Sigma}\cdot (-\mathbf{J}^{R}+\mathbf{T})+1}{\sqrt{[l\pm t]_{q}[(l\pm t)+3]_{q}}•}
\tag{8-9}
\end{equation}
Now, we can introduce our quantum fuzzy Ginsparg-Wilson system in instanton sector as follow:
\begin{equation}
\mathcal{A}_{qF}^{\pm}(\mathbf{T})=\langle\; \Gamma_{qF}^{\pm}(\mathbf{T}), \Gamma_{qF}^{'} : \Gamma_{qF}^{\pm^{^{2}}}(\mathbf{T})=\Gamma_{qF}^{'^{2}}=1,\quad\Gamma_{qF}^{\pm \dagger}(\mathbf{T})=\Gamma_{qF}^{\pm}(\mathbf{T}),\quad\Gamma_{qF}^{'\dagger}= \Gamma_{qF}^{'} \rangle.
\tag{8-10}
\end{equation}
Now, consider the following two elements constructed out of the generators $ \Gamma_{qF}(\mathbf{T}) $ and ${\Gamma_{qF}^{\prime}}  $ of the quantum fuzzy Ginsparg-Wilson algebra $ \mathcal{A}_{qF}(\mathbf{T}) $:
\begin{equation}
\begin{split}
\Gamma_{1qF}^{\pm}(\mathbf{T})=\Gamma_{qF}^{\pm}(\mathbf{T}) + {\Gamma_{qF}^{\prime}} \; , \qquad \qquad {(\Gamma_{1qF}})^{\dagger} = \Gamma_{1qF} ,\\
\Gamma_{2qF}^{\pm}(\mathbf{T}) = \Gamma_{qF}^{\pm}(\mathbf{T}) - {\Gamma_{qF}^{\prime}} \; , \qquad \qquad {(\Gamma_{2qF}})^{\dagger} = \Gamma_{2qF}.
\end{split}
\tag{8-11}
\end{equation}
Identifying $ \Gamma_{[(l\pm t)]_{q}} ^{L}$ and $ \Gamma_{[(l\pm t)]_{q}} ^{R}(T=0)$ with $ \Gamma $ and ${\Gamma^{\prime}} $, it is easy to compute $ \Gamma_{qF}^{\pm 1} $ and $ \Gamma_{qF}^{\pm 2} $. Now we can define q-deformed Dirac  and chirality operators as follow:
\begin{equation*}
D_{qF}^{\pm}(\mathbf{T})= \sqrt{[(l\pm t)]_{q}[(l\pm t)+3]_{q}}•\sqrt{[l]_{q}[l+3]_{q}}•\quad \Gamma_{qF}^{\pm1}(\mathbf{T})
\end{equation*}
\begin{equation}
\gamma_{qF}^{\pm}(\mathbf{T})= \sqrt{[(l\pm t)]_{q}[(l\pm t)+3]_{q}}•\sqrt{[l]_{q}[l+3]_{q}}•\quad \Gamma_{qF}^{\pm2}(\mathbf{T})
\tag{8-12}
\end{equation}
 which in the third commutative limi in (3-17) become:
\begin{equation}
 \lim_{l \to \infty, q \to 1} D_{qF}^{\pm }(\mathbf{T})=\pm (\mathbf{G} \cdot (\mathcal{G}+\mathbf{T})+2),\qquad  \lim_{l \to \infty, q\to 1}\gamma_{qF}^{\pm }(\mathbf{T}) =\pm \mathbf{G} \cdot \mathcal{\mathbf{x}}.
 \tag{8-13}
\end{equation}
These are the correct q-deformed Dirac and chirality operators on commutative four-sphere $ S^{4} $ in the instanton sector.
It is obvious that the Dirac operator (8-12) is $ \dagger $-invariant:
\begin{equation}
D_{qF}^{(\pm )^{\dagger}}(\mathbf{T})= D_{qF}^{(\pm )}(\mathbf{T}),
\tag{8-14}
\end{equation}
which we expect from commutative Dirac operator in instanton sector.

\section{Gauging the q-deformed fuzzy Dirac operator in instanton sector}
The derivation $\mathcal{J}_{i} $ dose not commute with the projectors $ \mathcal{P}_{qF}^{[l\pm t]_{q}} $ and then has no action on the modules $Mat(d_{l})\mathcal{P}_{qF}^{[l \pm t]_{q}}$. But $\mathcal{K}_{i}=q^{-1/2}\mathcal{J}_{i}+\dfrac{1}{•[2]_{q}}T_{i}$ does commute with $ \mathcal{P}_{qF}^{[l\pm t]_{q}}$. Here, $ \mathcal{K}_{i} $ has been considered as the total angular momentum.\\Now, we need to gauge $\mathcal{K}_{i} $. When $ T=0$, the gauge fields $ A_{i} $ were function of $ J_{i}^{L}$. Here, we consider $ A_{i}^{L} $ to be a functions of $ \mathbf{J}^{L}+\mathbf{T} $, because $ A_{i}^{L} $ dose not commute with $ \mathcal{P}_{qF}^{[l\pm t]_{q}} $. Let us introduce the covariant derivative as:
\begin{equation}
\nabla_{i}=\mathcal{K}_{i}+A_{i}^{L} .
\tag{9-1}
\end{equation}
In this case the limiting transversality of $ \mathbf{J}^{L}+\mathbf{T} $ can be guaranteed by imposing the condition:
\begin{equation*}
(\mathbf{J}^{L}+\mathbf{A}^{L}+\mathbf{T})\cdot(\mathbf{J}^{L}+\mathbf{A}^{L}+\mathbf{T})=(\mathbf{J}^{L}+\mathbf{T})\cdot(\mathbf{J}^{L}+\mathbf{T})=
\end{equation*}
\begin{equation}
\sum_{m=0,\pm1,\pm2} q^{m}(L^{L} + T)_{-m} (L^{L} +T)_{m}=[(l \pm t)]_{q}[(l\pm t)+3]_{q},
\tag{9-2}
\end{equation}
In the limit case we have:
\begin{equation}
\lim_{q \to 1}(\mathbf{J}^{L} + \mathbf{A}^{L}+\mathbf{T}) \cdot (\mathbf{J}^{L} + \mathbf{A}^{L}+\mathbf{T})=(l\pm t)((l\pm t)+3).
\tag{9-3}
\end{equation}
The expansion of (9-2) is:
\begin{equation}
q^{m} (J+T)_{-m}A_{m}+q^{m}A_{-m}(J+T)_{m}+q^{m}A_{-m}A_{m}=0.
\tag{9-4}
\end{equation}
When the parameter $ l $ tends to infinity, $ \dfrac{A_{i}^{L}}{l} $ tends to zero. For large $ l $, the (9-4) gives:
\begin{equation}
q^{m}x_{-m}^{L} A_{m}^{L} +q^{m} A_{-m}^{L} x_{m}^{L} +q^{m} \dfrac{A_{-m}^{L} A_{m}^{L}}{l} = 0.
\tag{9-5}
\end{equation}
$ A_{i}^{L}$is to remain bounded as $ l $ tends to infinity. Also, in this limit $ x_{i}^{L} $ tends to $ \hat{x}_{i}$, the unit normal to the $ S^{4} $ at $ \hat{\mathbf{x}} $. So in the limiting case, if $ A_{i}^{L} $ tends to $ a_{i} $ then $ \hat{\mathbf{x}} \cdot \mathbf{a}=0 $.\\
Now, we can construct the gauged q-deformed fuzzy Ginsparg-Wilson system in instanton sector and its corresponding Dirac and chirality operators as follow: 
In this case the quantum left projectors are given by
\begin{equation}
\mathcal{P}_{[(l\pm t)]_{q}}^{L}(\mathbf{T}, \mathbf{A}^{L})= \dfrac{1}{[2]_{q}•}[1\pm \dfrac{\mathbf{\Sigma}\cdot (\mathbf{J}^{L}+ \mathbf{T}+ \mathbf{A}^{L})+1}{\sqrt{[(l\pm t)]_{q}[(l\pm t)+3]_{q}}•}],\quad \mathcal{P}_{[(l\pm t)]_{q}}^{L\dagger}= \mathcal{P}_{[(l\pm t)]_{q}}^{L},
\tag{9-6}
\end{equation}
and the corresponding involutions have the form
 \begin{equation}
\Gamma_{qF}^{\pm L}(\mathbf{T,A^{L}})=[2]_{q} \mathcal{P}_{[(l\pm t)]_{q}}^{L}-1=\pm \dfrac{\mathbf{\Sigma}\cdot (\mathbf{J}^{L}+ \mathbf{T}+ \mathbf{A}^{L})+1}{\sqrt{[(l\pm t)]_{q}[(l\pm t)+3]_{q}}•},
\tag{9-7}
\end{equation}
We choose $ \Gamma_{qF}^{'} $ as in (7-13).
The quantum projectors $\mathcal{P}_{[(l\pm t)]_{q}}^{R} $ coupling the right momentum, instanton and spin $ \dfrac{1}{2} $ to its maximum and minimum values $ (l\pm t)\pm \dfrac{1}{2} $, respectively are obtained by changing $ J_{i}^{L} $ to $ -J_{i}^{R}$ in the above expression
 \begin{equation}
\mathcal{P}_{[(l\pm t)]_{q}}^{R}(\mathbf{T}, \mathbf{A}^{R})= \dfrac{1}{[2]_{q}•}[1\pm \dfrac{\mathbf{\Sigma}\cdot (-\mathbf{J}^{R}+ \mathbf{T}+ \mathbf{A}^{R})+1}{\sqrt{[(l\pm t)]_{q}[(l\pm t)+3]_{q}}•}],\quad \mathcal{P}_{[(l\pm t)]_{q}}^{R\dagger}= \mathcal{P}_{[(l\pm t)]_{q}}^{R},
\tag{9-8}
\end{equation}
and the corresponding idempotents can be given by
\begin{equation}
\Gamma_{qF}^{\pm R}(\mathbf{T,A^{R}})=[2]_{q} \mathcal{P}_{[(l\pm t)]_{q}}^{R}-1=\pm \dfrac{\mathbf{\Sigma}\cdot (-\mathbf{J}^{R}+ \mathbf{T}+ \mathbf{A}^{R})+1}{\sqrt{[(l\pm t)]_{q}[(l\pm t)+3]_{q}}•},
\tag{9-9}
\end{equation}
Now, we can construct the gauged quantum  fuzzy Ginsparg-Wilson system in instanton sector and its corresponding quantum fuzzy Dirac and chirality operators as follow: 
\begin{equation*}
\mathcal{A}_{qF}^{\pm}( \mathbf{T},\mathbf{A}^{L})= \langle \Gamma_{qF}^{\pm}(\mathbf{T,A^{L}}), \Gamma_{qF}^{'} : \Gamma_{qF}^{\pm^{^{2}}}(\mathbf{T,A^{L}})=\Gamma_{qF}^{'^{2}}=1,
\end{equation*}
\begin{equation}
\Gamma_{qF}^{\pm \dagger}=\Gamma_{qF}^{\pm} ,\quad\Gamma_{qF}^{'\dagger}= \Gamma_{qF}^{'} \rangle.
\tag{9-10}
\end{equation}
Now, consider the following two elements constructed out of the generators $ \Gamma_{qF}(\mathbf{T,A^{L}}) $ and ${\Gamma_{qF}^{\prime}}  $ of the quantum fuzzy Ginsparg-Wilson algebra $ \mathcal{A}_{qF}(\mathbf{T,A^{L}}) $:
\begin{equation}
\begin{split}
\Gamma_{qF}^{\pm1}(\mathbf{T,A^{L}})=\Gamma_{qF}^{\pm}(\mathbf{T,A^{L}}) + {\Gamma_{qF}^{\prime}} \; , \qquad \qquad {(\Gamma_{1qF}})^{\dagger} = \Gamma_{1qF},\\
\Gamma_{qF}^{\pm2}(\mathbf{T,A^{L}}) =\Gamma_{qF}^{\pm}(\mathbf{T,A^{L}}) - {\Gamma_{qF}^{\prime}} \; , \qquad \qquad {(\Gamma_{2qF}})^{\dagger} = \Gamma_{2qF}.
\end{split}
\tag{9-11}
\end{equation}
Identifying $ \Gamma_{[(l\pm t)]_{q}} ^{L}$ and $ \Gamma_{[(l\pm t)]_{q}} ^{R}(T=0)$ with $ \Gamma $ and ${\Gamma^{\prime}} $, it is easy to compute $ \Gamma_{qF}^{\pm 1} $ and $ \Gamma_{qF}^{\pm 2} $. Now we can define q-deformed Dirac  and chirality operators as
\begin{equation*}
D_{qF}^{\pm}(\mathbf{T,A^{L}})= \sqrt{[(l\pm t)]_{q}[(l\pm t)+3]_{q}}•\sqrt{[l]_{q}[l+3]_{q}}\quad•\Gamma_{1qF}^{\pm}(\mathbf{T},\mathbf{A}^{L})
\end{equation*}
\begin{equation}
\gamma_{qF}^{\pm}(\mathbf{T,A^{L}})= \sqrt{[(l\pm t)]_{q}[(l\pm t)+3]_{q}}•\sqrt{[l]_{q}[l+3]_{q}}•\quad \Gamma_{2qF}^{\pm}(\mathbf{T},\mathbf{A}^{L})
\tag{9-12}
\end{equation}
 which in the third commutative limit in (3-17) become:
\begin{equation}
 \lim_{l \to \infty, q\to 1} D_{qF}^{\pm }(\mathbf{T}, \mathbf{A}^{L})=\pm(\mathbf{G} \cdot (\mathcal{G}+ \mathbf{T}+ \mathbf{A}^{L})+2),\qquad  \lim_{l \to \infty, q\to 1}\gamma_{qF}^{\pm }(\mathbf{T}, \mathbf{A}^{L}) =\pm \mathbf{G} \cdot \mathcal{\mathbf{x}},
 \tag{9-13}
\end{equation}
which we expect from commutative gauged Dirac and chirality operators in instanton sector on $ S^{4} $.

\section{Conclusion}
\textbf{}
In this paper, using the q-deformed projectors and idempotents of the finitely generated q-deformed projective $ \mathcal{A}(S_{qF}^{4})  -$module it has been constructed the generators of the q-deformed gauged fuzzy Ginsparg-Wilson algebra in instanton and no-instanton sector. It has been constructed q-deformed gauged fuzzy Dirac operator in instanton and no-instanton sector using the q-deformed fuzzy GW algebra. The importance of this Dirac operator is that it has correct commutative limit.

\end{document}